\documentclass[aps,prl,twocolumn,groupedaddress]{revtex4-1}
\usepackage{graphicx}
\usepackage{dcolumn}
\usepackage{float}
\usepackage{verbatim}
\usepackage{xcolor}
\usepackage{amsmath}
\usepackage{xcolor, colortbl}

\begin{document}


\title{Extending the Nanbu Collision Algorithm to Non-Spitzer Systems and Application to Laser Heating and Damage} 



\author{A. M. Russell and D. W. Schumacher}
\affiliation{Department of Physics, The Ohio State University, 191 W. Woodruff Ave, Columbus, OH 43210}


\date{\today}

\begin{abstract}
We have generalized the Nanbu collision algorithm to accommodate arbitrary collision rates, enabling accurate kinetic modeling of short range particle interactions in non-Spitzer systems. With this extension, we explore the effect of different collision models on the simulation of how ultra-intense lasers first begin to heat a target. The effect of collisions on plasma evolution is crucial for treating particle slowing, energy transport, and thermalization.
The widely used Nanbu collision algorithm provides a fast and computationally efficient method to include the effects of collisions between charged particles in kinetic simulations without requiring that the particles already be in local thermal equilibrium. However, it is ``hardwired" to use Spitzer collision rates appropriate for hot, relatively dilute plasmas. This restriction prevents the Nanbu collision algorithm from accurately describing the initial heating of a cold target, a key problem for the study of laser damage or the generation of the warm dense matter state. We describe our approach for modifying the Nanbu collision algorithm and demonstrate the improved accuracy for copper targets.
\end{abstract}


\maketitle 

\section{Introduction}
Ultra-intense laser excitation of cold, solid density targets drives rapid heating and highly non-equilibrium distributions of electrons in a process that is still not well understood. Elucidating these dynamics is a crucial step for the study of laser generated warm dense matter and laser induced surface modification and damage \cite{warm_dense_matter_1,warm_dense_matter_2, surface_modification_1, surface_modification_2, optical_ablation, Rob}. In particular, a better understanding is important for optimizing applications of ultrashort (sub-ps), ultra-intense ($> 10^{12}$ $\mathrm{W/cm^{2}}$) laser driven surface modification which includes laser surgery, laser machining and characterization, creation of hydrophobic surfaces, solar cell enhancement and many others \cite{eye_surgery, laser_machining_1, laser_machining_2, hydrophobic_1, hydrophobic_2, solar_cell, lithography}. For laser pulses achieving relativistic intensities (approximately $10^{18} \; \mathrm{W/cm^{2}}$ for near-infrared lasers), this physics is germane as well since the initial heating that occurs on the leading edge of the laser pulse at much lower intensities determines the nature of the surface once the peak of the pulse arrives. Despite the wide-ranging importance of laser heating of cold targets, it remains an open theoretical problem and computational methods are widely used. An important approach to modeling in this regime is the kinetic simulation, such as the particle in cell (PIC) method, because non-thermal electron distributions can dominate the evolution of the target. Within these simulations the randomization of motion due to collisions must be accounted for since they determine the energy transport and eventual thermalization.

An important class of collisions is that between two charged particles where long range potentials drive scattering processes that are not well described as point-like interactions. The prevalence of plasmas where these collisions dominate has prompted the development of a multitude of schemes for the numerical simulation of collisional plasma behavior, for example, those that operate at the level of the particle distribution function and those that operate at the discrete particle level. A collision mechanism is usually parameterized using the temperature and density dependent collision rate, however this is most naturally applied in equations of motion for the distribution function. In collisions between two particles there is no rate as such; instead, the rate characterizes time or ensemble averages. The Binary Collision Model (BCM), a Monte Carlo method pioneered by Takizuka and Abe \cite{Takizuka} and improved upon by Nanbu\cite{Nanbu}, connects the scattering rate for Spitzer collisions \cite{Spitzer} to the distribution of scattering angles that would be observed for an ensemble of particle-pairs. Takizuka and Abe introduced an efficient algorithm for sampling this distribution and modifying the particle trajectories accordingly and Nanbu was able to relax the constraint of resolving the collision rate by treating the effect of a large number of small-angle collisions. One powerful advantage of the BCM is its ability to treat non-thermal distributions - a necessary capability for modeling ultra-intense laser-plasma interactions. Although identification of a temperature scale for the background particles is needed to establish the collision rate, the BCM does not explicitly require the scattering particles to conform to any particular distribution. However, the ``hardwiring" of the BCM to the Spitzer collision rate means the BCM cannot be used for realistic modeling of the initial heating of a target by an ultrashort, ultra-intense laser. The Spitzer rate diverges for low temperatures, causing the dephasing and randomization of motion of the heated electrons to be greatly exaggerated. This in turn results in inaccurate modeling of key parameters such as the reflectivity and skin depth which determine the amount and physical extent of the heating. Once the target becomes sufficiently hot, T = 10 eV or more, this error becomes small, but this may not occur until the end of the pulse if at all for the pulse parameters used for laser surface modification experiments. For pulses achieving relativistic intensities, it is still an open question how important the initial heating during the leading edge of the pulse is. Recent results studying laser driven ion acceleration suggest it can play a key role, for example \cite{GSI}.  Thus, a general approach is required that can accurately treat non-thermal particle distributions with non-Spitzer collision rates. We show that the BCM can be readily generalized to provide just this approach. We demonstrate its application to the ultrafast, ultra-intense heating of a copper target modeled using the PIC code LSP \cite{LSP}, comparing the results for different choices of collision rate and collision algorithm. In particular, we examine the effect on the resulting electron distribution and the treatment of diffusive and ballistic motion of the heated electrons.

\section{Theory}

The collisionality of a system is typically characterized by a collision frequency $\nu_{\parallel}$ defined by the formula $\frac{d \mathbf{v}}{d t} = \nu_{\parallel} \mathbf{v} $ that indicates the rate at which a particle's velocity $\mathbf{v}$ is attenuated. However, collisional behavior can be equally well described by an angular deflection frequency defined as $\frac{d \chi^{2} }{d t} = \nu_{\perp} \chi^{2}$, where $\chi^{2}$ is the squared angular deviation from the initial trajectory. Under the condition that consecutive collisions are uncorrelated, one can show $\nu_{\perp} = 2\nu_{\parallel}$\cite{Jackson}, providing a connection between processes involving frictional slowing and angular diffusion. With this, we can deduce 
\begin{equation}
\label{main_equation}
\frac{\langle \theta^{2} \rangle}{2} = \nu_{\parallel}\delta t, 
\end{equation}
where $\theta$ is the deflection induced by a single collision and $\delta t$ is the time during which the deflection occurs. The average $\langle \rangle$ is taken over time, or, equivalently for ergodic systems, the velocity distribution of the particles. 

While the average angular diffusion obeys this equation, any particular collision between two particles will be determined by the velocity difference and impact parameter. The second of these quantities may be integrated out to yield a velocity dependent scattering parameter $s\left(\mid \mathbf{v_{1}} - \mathbf{v_{2}}\mid\right) = s\left(\mathbf{u}\right) =  \theta^{2}/2$. This parameter determines how the average angular deflection per unit time depends on the relative speed of the colliding particles. Once the general form of $s$ is known, the BCM procedure can be applied. Briefly, the set of colliding particles is randomized and divided into pairs such that each particle only collides with its partner. The relative speeds of each pair of particles is used to determine $s$, from which the scattering angle can be found via the formalism developed by Nanbu. By applying the scattering angle to the relative velocity of the particles, they experience an energy and momentum conserving collision.

We now consider the case of an arbitrary distribution of heated particles scattering against a thermal background characterized by a rate $\nu_{\parallel}$ which need not be Spitzer. If we average $s$ over the background distribution, we obtain the equation
\begin{equation}
\langle s\left(\mathbf{u}\right) \rangle = \nu_{\parallel}\delta t.
\end{equation} 
For a Maxwell-Boltzmann distribution this may be written as
\begin{align}
\label{mb_main}
\langle s\left(\textbf{u}\right) \rangle =& \left(\frac{m_{\alpha}}{2 \pi k_{B} T_{\alpha}}\right)^{3/2}
\left(\frac{m_{\beta}}{2 \pi k_{B} T_{\beta}}\right)^{3/2}\nonumber \\ \times& \int d^{3}v_{\alpha}\int d^{3} v_{\beta} e^{-\frac{m_{\alpha} v_{\alpha}^{2}}{2 k_{B} T_{\alpha}}}e^{-\frac{m_{\beta} v_{\beta}^{2}}{2 k_{B} T_{\beta}}}s\left(\mid \mathbf{v_{\alpha}} - \mathbf{v_{\beta}}\mid\right) \nonumber \\
=& \frac{4}{\sqrt{\pi}}\int_{0}^{\infty}dx\; e^{-x^2}x^{2}\; s\left(\kappa x\right) = \nu_{\parallel}\delta t,
\end{align}
where 
\begin{equation}
\kappa = \sqrt{{2 k_{B}\left(\frac{T_{\alpha}}{m_{\alpha}} + \frac{T_{\beta}}{m_{\beta}}\right)}},
\end{equation}
$k_{B}$ is Boltzmann's constant, $\beta$ and $\alpha$ denote, respectively, the deflected and background particle, and T and m are the temperatures and masses of the corresponding particles.

The goal is then to find the form of the relative velocity dependent scattering parameter $s$ such that individual collisions are accurately modeled and Eq. \ref{mb_main} is satisfied. We now present a procedure for doing this by a straightforward modification of the result for Spitzer collisions. For that case, $s$ may be written as 
\begin{equation}
s = \frac{\ln \Lambda}{4\pi}\left(\frac{q_{\alpha}q_{\beta}}{\epsilon_{0}m_{\alpha}}\right)^{2}\frac{1}{u^{3}}n_{\alpha}\delta t,
\end{equation}
where $\ln \Lambda$ is the Coulomb logarithm, $q$ represents charge and $n$ represents number density. For energetic particles undergoing glancing, low angle collisions, this equation is valid. However, the divergence at zero velocity and consequent divergence of the integral in Eq. \ref{mb_main} invalidates the definition of a collision rate, so $s$ must first be modified appropriately. A reasonable approximation is to model the scattering as isotropic below a velocity cutoff $u_{0}$, which corresponds to a modified scattering parameter
\begin{equation}
\label{s0}
s\left(u\right) = \frac{\Sigma}{4\pi}\left(\frac{q_{\alpha}q_{\beta}}{\epsilon_{0}m_{\alpha}}\right)^{2}\frac{n_{\alpha}}{u_{0}^{3}}g\left(u\right)\delta t,
\end{equation}
where $g\left(u\right)$ is 
\begin{equation}
\label{gfunc}
g\left(u\right) = \left[\Theta\left(u - u_{0}\right)\left(\frac{u_{0}}{u}\right)^{3} + \Theta\left(u_{0} - u\right)\left(\frac{u}{u_{0}}\right)\right],
\end{equation}
and $\Theta$ is the Heaviside-theta function. To generalize the result for non-Spitzer collision models, we have replaced the usual Coulomb logarithm with $\Sigma$ to serve as a generalized scaling parameter which must be chosen to satisfy Eq. \ref{mb_main}. The velocity cutoff $u_{0}$ is chosen to be that at which a glancing collision at average interparticle distance yields a $45^{\circ}$ deflection:
\begin{equation}
u_{0} = \frac{e}{\sqrt{2\pi \epsilon_{0} \mu}}n_{\alpha}^{1/6}, 
\end{equation}
where $\mu$ is the reduced mass. This leads us to a scaling parameter of
\begin{equation}
\Sigma = \sqrt{\frac{2}{\pi}}\frac{1}{\langle g\left(u\right)\rangle}\frac{\nu_{\parallel} }{\omega_{p}},
\end{equation}
where the expectation value of the scattering function $g\left(u\right)$ can be found to be
\begin{equation}
\nonumber
\langle g\left(u\right)\rangle = \frac{2}{\sqrt{\pi} \chi} \left[\left(1 + \chi^{4}\Gamma\left(0, \chi^{2}\right)\right) - e^{-\chi^{2}}\left(1 + \chi^{2}\right)\right].
\end{equation}
Here $\chi = u_{0}\sqrt{m/k_{B}T}/2$ and $\Gamma$ is the incomplete gamma function.

To summarize, given a collision rate $\nu_{\parallel}$, the generalized scaling parameter $\Sigma$ is found from Eq. 9 and the scattering parameter $s$ from Eq. 6. The BCM can then be implemented in the usual way. It should be noted that this form of $\langle g\left(u\right) \rangle$ is dependent upon the choice of $s$, and may need to be modified appropriately if individual collisions are not properly described by Eq. \ref{gfunc}. Also, although sensitivity to time step has been assessed for standard Nanbu collisions \cite{Nanbu_vs_TA}, any custom collision rates may have an impact on the numerical convergence and should be independently tested.

\section{Validity and Accuracy of the Method}

The purpose of our modification to the Nanbu collision algorithm is to allow for the accurate modeling of properties like skin depth and absorption which will vary rapidly as a function of space and time during ultrafast, ultra-intense laser heating. This requires the use of collision rates appropriate for initially cold, dense targets that undergo non-thermal heating with steep temperature and density gradients present for at least part of the heating process. With this in mind, we perform two different checks of our approach and provide an illustration of the resulting nonthermal behavior using a test system of a thin copper target. Accordingly, this section covers:

\begin{enumerate}
	\item A test of the satisfaction of Eq. \ref{main_equation}.
	\item A test of the agreement of the simulated low fluence, room temperature absorption with a theoretical estimate.
	\item An examination of the prevalence of nonthermal behavior in the high fluence regime.
\end{enumerate}

As a replacement for Spitzer rates, which are clearly invalid in this regime, we first select more appropriate electron-ion and electron-electron collision rates. The LMD model \cite{LM_Rates, LMD_Rates} describes electron-ion collisions across a wide range of temperatures and densities and is well established for particle based simulations. Electron-electron collisions are well understood in metals at high and low temperatures, as these correspond to the Spitzer and solid-state regimes. The lack of theory concerning the intermediate regime prompted Colombier et. al \cite{colombier} to approximate the associated collision rates with a cubic spline, which is what we use for our simulation of copper. These electron-electron and electron-ion rates will collectively be referred to as copper rates. Finally, in this section, we also compare the BCM to the Jones model \cite{Jones}, which is a grid based model that assumes that each species conforms to a Maxwell-Boltzmann distribution. All simulations were performed using the PIC code LSP \cite{LSP} in 2D3V geometry. LSP incorporates both the BCM and Jones models and was modified to use the scheme introduced in this work.

\subsection{Adherence to Specified Scattering Rate}

\begin{figure}
	\centering
	\includegraphics[width=0.95\linewidth]{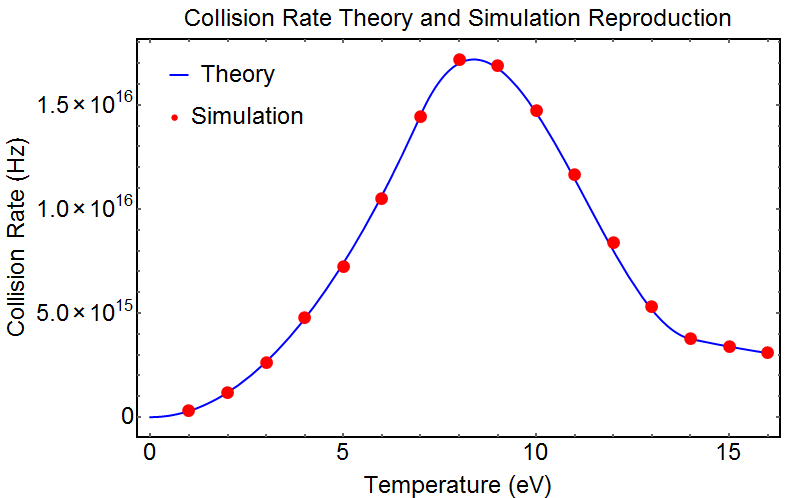}
	\caption{Shown are the Colombier electron-electron collision rates for solid copper (blue line) and those which were found from the simulation (red dots). The standard deviation for each data point is less than the size of the dots.}
	\label{fig:Colombier_Angular_Test}
\end{figure}

Using our modified BCM approach, the validity of Eq. \ref{main_equation} was checked for electron-electron collisions with the Colombier rates by initializing a solid density (8.5$\times 10^{22}$ $\mathrm{cm}^{-3}$), singly ionized copper block to a homogeneous temperature and sampling the angular deflection undergone by the electrons in it during 20,000 collisions. In this simulation the cell size was $3.25\; \mathrm{nm} \times 3.25\; \mathrm{nm}$ with 900 electrons in each cell and with a time step of $\tau = 4.6\times 10^{-18}$. The ions were approximated by a homogeneous positive background. Shown in Fig. $\,\!$\ref{fig:Colombier_Angular_Test} are the theoretical collision rates for a range of temperatures and the collision rates reproduced by the simulation. The excellent agreement shows that at the most basic level, binary collisions do reproduce the desired collision rate. 

\subsection{Cold target absorption}

We test the absorption of a room temperature copper block by subjecting it to a low-intensity beam. The dynamics in this scenario are well described by the Fresnel equations from linear optics, where the absorption at normal incidence from vacuum to a material of refractive index n is
\begin{equation}
A = 1 - \left|\frac{1 - n}{1 + n}\right|^{2}.
\end{equation}

Moreover, for a specified collision rate, Drude\cite{Drude} theory dictates that the index of refraction is 
\begin{equation}
n = \left(1 - \frac{\omega_{p}^{2}}{\omega^{2} + i \omega\nu_{\parallel}}\right)^{1/2}, 
\end{equation}
where $\omega$ is the frequency of incident light and $\omega_{p}$ is the plasma frequency. The combination of these two theories adequately describes the situation in which ions and electrons are represented by appropriately charged monopoles, and is therefore used as a benchmark for the absorptivities produced via simulation. Those absorptivities listed in Table \ref{table:Absorption_Table} from simulation results correspond to 800 nm light in the form of a Gaussian beam focused to the surface with intensity full width half maximum (FWHM) pulse length 50 fs, intensity FWHM spot size 0.59 $\mathrm{\mu m}$, and peak fluence $1.66 \;\textrm{mJ}/\textrm{cm}^{2}$. The cell resolution was 5 nm in the longitudinal direction and 10 nm in the transverse. By recording the net Poynting flux into and out of the simulation, the absorbed light was calculated. 

\begin{table}
	\begin{ruledtabular}
	\begin{tabular}{l | l | l}
		\multicolumn{3}{ c }{Absorptivities for Various Models and Rates} \\
		\hline\hline
		e-e rates/model & e-i rates/model & absorption \% \\
		\hline
		Capped Spitzer/Binary & Capped Spitzer/Binary & 89.6 \\
		Colombier/Jones & LMD/Jones & 3.3 \\
		Colombier/Binary & LMD/Binary & 2.1 \\
		\hline
		\multicolumn{2}{c|}{Drude-Lorentz Prediction} & 1.7
	\end{tabular}
	\end{ruledtabular}
	\caption[Absorptions]{The following absorptions were calculated using the rates and models labeled. The capped Spitzer rates were restricted to frequencies $\nu < \nu_{cap} = 10^{17} \mathrm{Hz}$.}
	\label{table:Absorption_Table}
\end{table}

Without the ability to insert the correct collision rates into the BCM, the absorption can only be determined via Spitzer physics. For a solid target at room temperature this presents multiple issues.  Strictly speaking, Spitzer rates are undefined in this regime due to the Coulomb logarithm $\ln \Lambda$ being negative. While setting a minimum $\ln \Lambda$ fixes this problem, the resulting collision time is several orders of magnitude lower than the maximum time step imposed by the Courant limit and the necessary spatial resolution, leading to collisions that cannot be calculated in a numerically converged way in a reasonable amount of time. For this reason we calculated a lower bound on the absorption resulting from Spitzer collisions by setting $\ln \Lambda = 2$ and capping the collision frequency at $\nu_{cap} = 10^{17} \mathrm{Hz}$, which approached the highest numerically converged frequency accessible with our simulation parameters. Even with these modifications in place, the simulated absorption exceeded the theoretical prediction by over an order of magnitude.

The utilization of the Jones model and inclusion of the copper rates provided a reasonable answer differing by a factor of two from theory but still overestimated the absorption by excessively increasing the temperature of the electrons. The closest match to the Drude-Lorentz prediction was produced by modeling the collisionality with copper rates and binary collisions, giving an error of only 24\%, significantly better than the other approaches. 

\begin{figure}
	\centering
	\includegraphics[width=.95\linewidth]{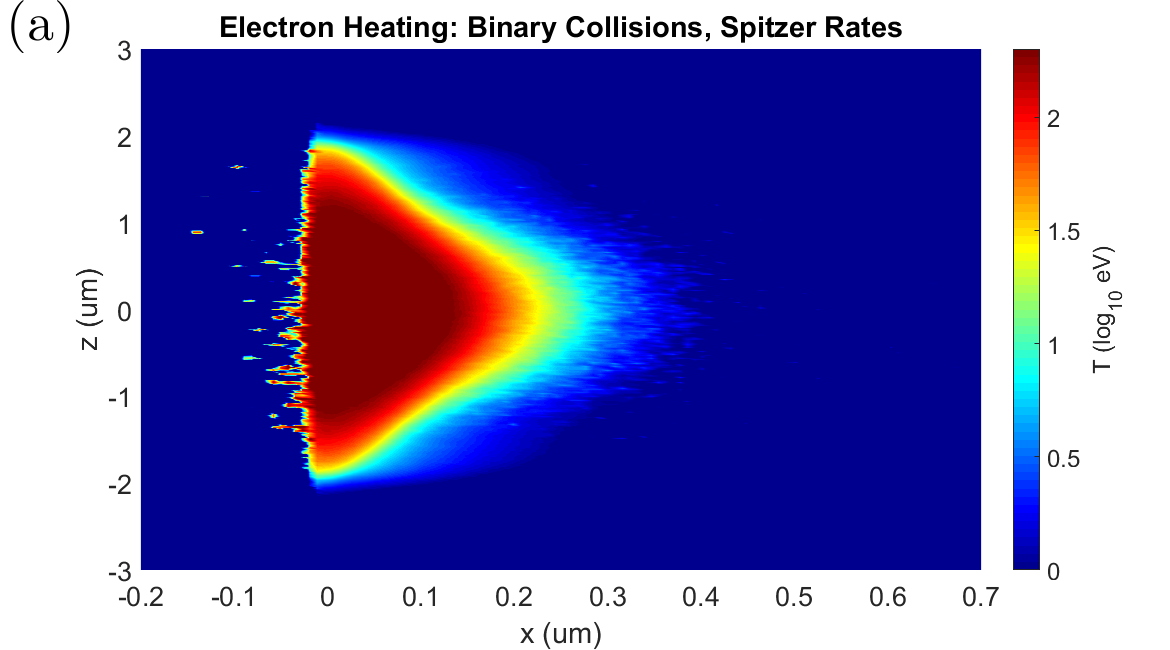}
	\includegraphics[width=.95\linewidth]{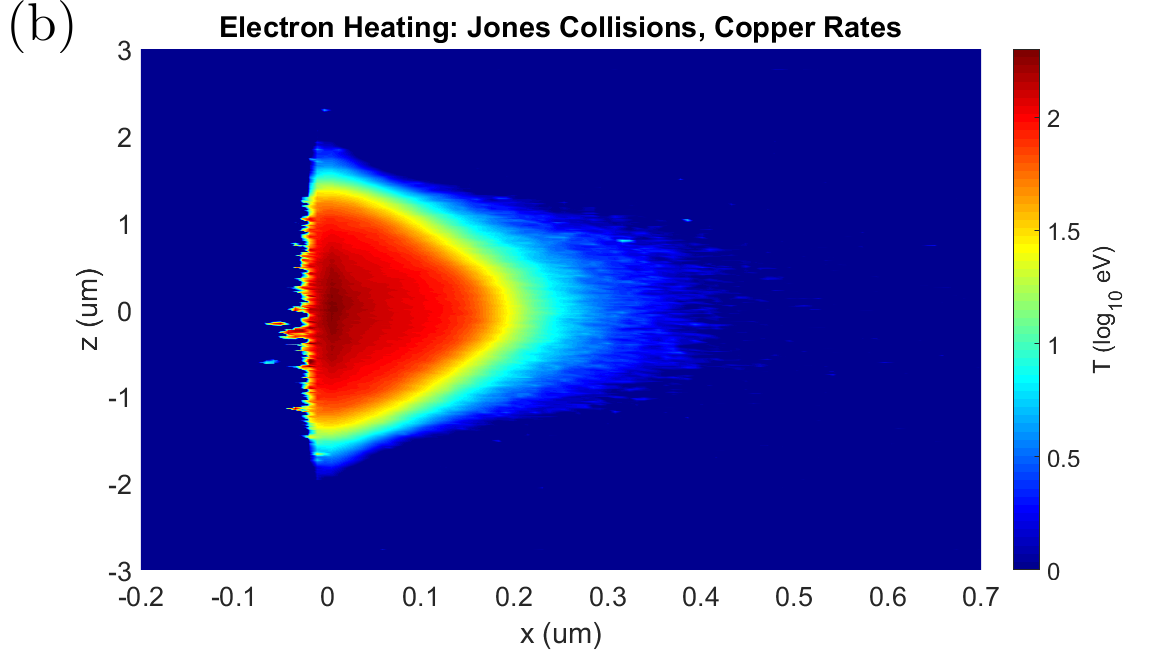}
	\includegraphics[width=.95\linewidth]{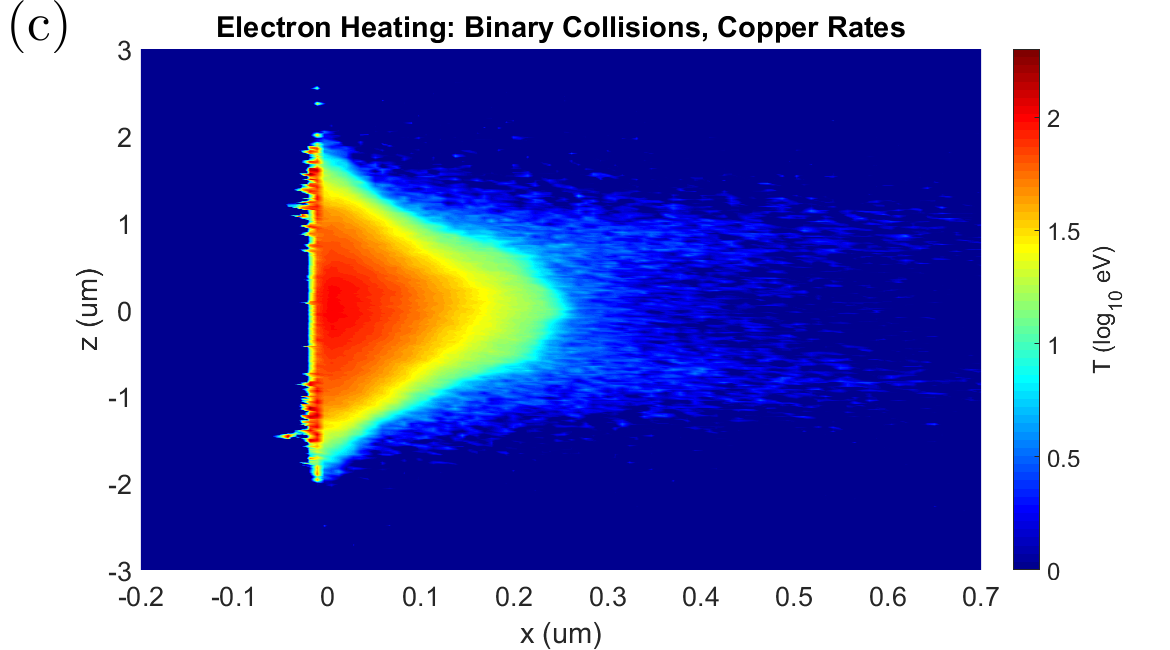}	
	\includegraphics[width=.95\linewidth]{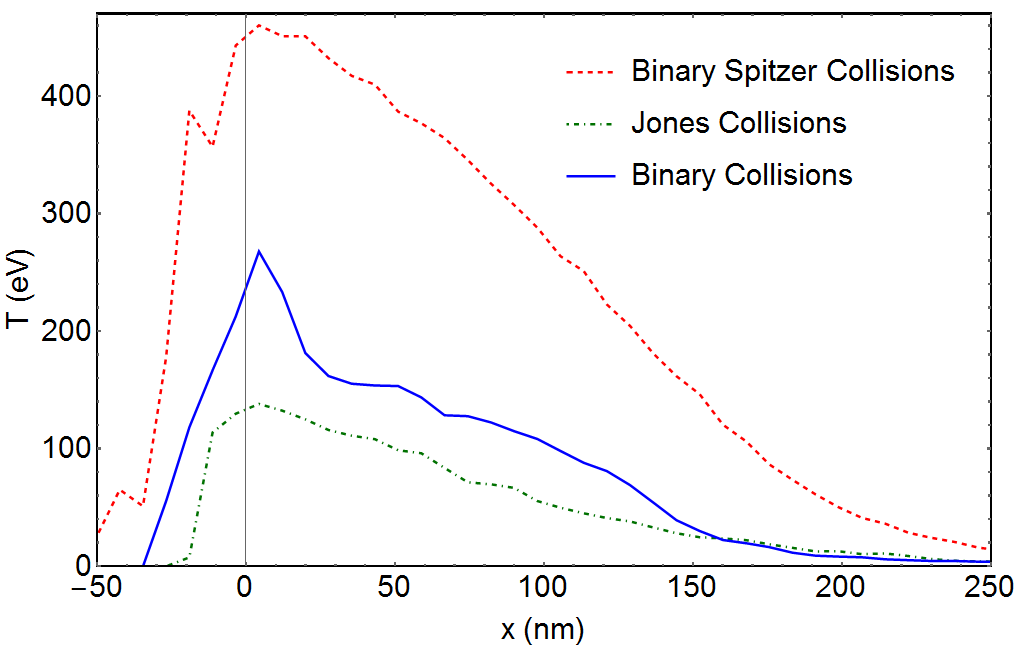}
	\caption[short cap]{Electron temperature after the laser has left the grid for the indicated combinations of collision algorithm and collision rate. (a-c) 2D spatial distributions shown using a log color scale. The laser enters the simulation from the left, hitting the target surface along x = 0. (d) Electron temperature along the line z = 0. The differences in collision model and rate result in differing heating patterns. Note the diffuse electron population outside the target at this time due to the pressure inside the target and ballistic motion.}
	\label{fig:high_fluence_heating}
\end{figure}

\begin{figure}
	\centering
	\includegraphics[width=1\linewidth]{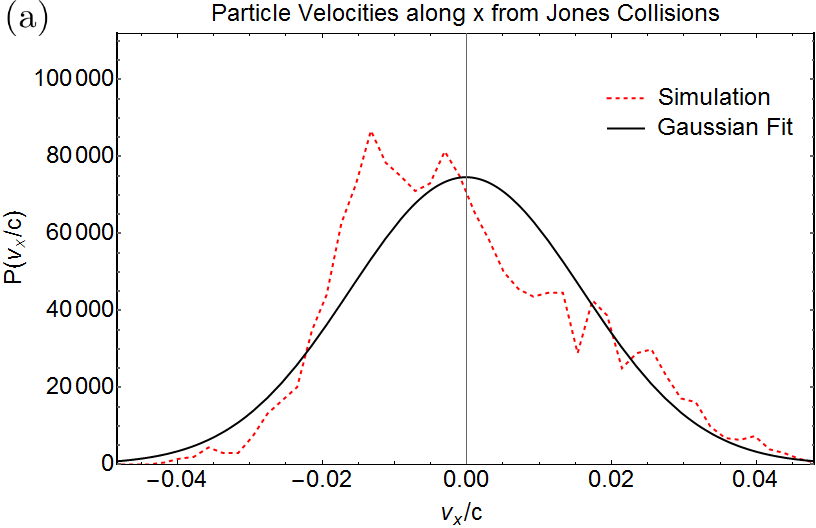}
	\includegraphics[width=1\linewidth]{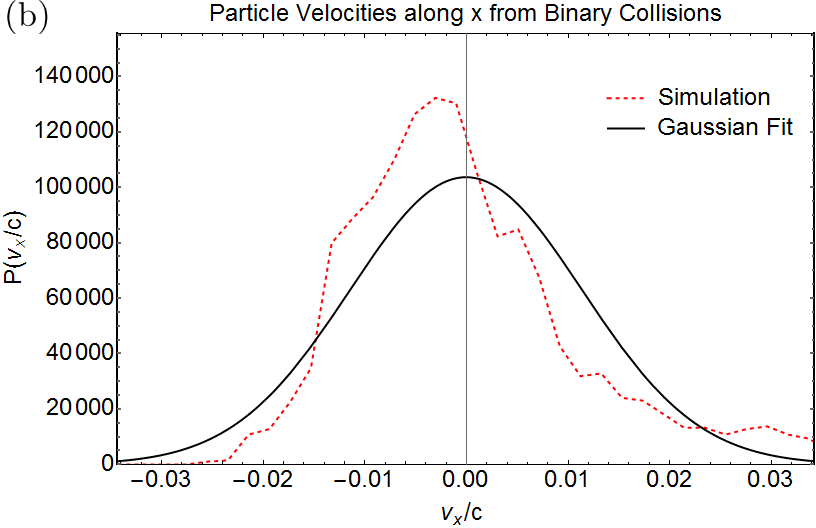}
	\caption{Longitudinal electron velocity distributions near the time of peak excitation (red) and Gaussian fit (blue). Copper rates were applied using (a) the Jones algorithm and (b) the modified binary collisions algorithm. Binary collisions show stronger non-thermal behavior in the form of increased skewness. The distributions were calculated by grouping electron macroparticles from the region defined by $60 \;\mathrm{nm} < x < 70 \; \mathrm{nm}$ and $-10 \;\mathrm{nm} < z < 10 \;\mathrm{nm}$ at three quarters of the way through the laser pulse. The Gaussian fits were found based on the standard deviations and means of those data sets. In order to remove the effect of temperature difference from the visual comparison of the plots, the choice of horizontal and vertical extent was scaled to the corresponding Gaussian fit.}
	\label{fig:Collisions_Hist}
\end{figure}

\subsection{Nonthermal behavior}

While reproducing the appropriate absorption is necessary for accurately modeling the laser-induced heating of a target, a key advantage of using particle based collisions is the ability to resolve single particle effects not described by a thermal distribution. The prominence of these effects was tested by simulating the laser heating induced by a single 40 fs FWHM, 780 nm wavelength pulse with a spotsize of  $0.59\; \mathrm{\mu m}$ FWHM and intensity of $10^{17}\; \textnormal{W}/\textnormal{cm}^{2}$ at normal incidence on a 1 $\mathrm{\mu m}$ thick copper block. The simulation was run with a cell resolution of $1/128\;\mathrm{\mu m}$ in the longitudinal direction, $1/64\; \mathrm{\mu m}$ in the transverse direction and a time step of $\tau = 7\times10^{-18} \mathrm{s}$. In each cell the number of electron macroparticles was 3600 and the number of ion macroparticles was 900. Although the laser intensity used was sufficient to induce ionization, the generation of additional electrons and ion species was suppressed so as not to obfuscate the effects of the collisionality. Three different permutations of models and rates were analyzed: binary collisions with uncapped Spitzer rates and copper rates and Jones collisions with copper rates. 

The electron temperature after the laser has hit the target and left the grid is shown in Fig. 2 for the cases described above. Fig. 2(a-c) shows the spatial temperature distribution in which clear differences in the temperature and shape of the heated area are visible. Even though the copper rates approach Spitzer rates as the temperature increases, Fig. 2(a) shows that using Spitzer rates for the entirety of the simulation leads to a dramatic increase in heating, consistent with the absorption calculated in Table \ref{table:Absorption_Table}. The choice of copper rates but with different collision models as in Fig. 2(b-c) produces heating profiles with similar temperatures but significantly different heating profile shapes. Jones collisions suppress ballistic electrons, producing a hemispherical heating pattern that expands in an essentially isotropic manner, while binary collisions more accurately simulate individual electron dynamics and create a heating pattern that more closely mirrors the spatial variation of the laser intensity. Additionally, binary collisions generate a trail of energy deposition leading to the back of the target. The Jones model inhibits this feature by virtue of not being able to describe a small collection of energetic particles moving through an otherwise cool background. Specifically, if a single hot electron at temperature $T_{e}$ is moving through a background of cold ions at temperature $T_{i}$, the resulting velocity damping of the electron will be reduced by a factor of $\left(\left(T_{e}/T_{i}\right) + N_{e}\right)$, where $N_{e}$ is the number of electrons in the cell through which the ballistic is traveling. This induces multiple effects; the deposition of energy into the front, center, and back of the target will be altered along with the collisionality in that region, and sufficiently energetic electrons can incite impact ionization in regions beyond their maximum penetration depth. 

The differences in heating can be seen more quantitatively by observing a cut along the center of the material as in Fig. 2(d). When using Spitzer collisions the peak temperature differs by as much as a factor of four and the total absorbed energy by a factor of five. Not only is this in itself a significant deviation, but it results in a collision rate varying by nearly an order of magnitude, a difference that will impact further heating in simulations where the peak of the laser has yet to reach the material. 

Finally, we note that the ion temperature is little changed at the time shown in the figure and electron-ion equilibration will not be achieved under these conditions for 10’s of ps. The energy available for surface modification and damage is almost entirely contained in the electronic system at this time, and its scale and spatial extent will determine the evolution of the target.

Additional insight can be gained from the distribution of electron velocities in the laser propagation direction (x) shown in Fig. \ref{fig:Collisions_Hist}. We see that when compared to the histogram produced by binary collisions, Jones collisions dampen the velocities of particles propagating parallel to the laser. While this does not completely eliminate the skew of the distribution, it forces the particle velocities to conform unrealistically closely to a Gaussian and thereby decreases the weight of the tail. This variation in particle distribution between models drives the macroscopic properties of the system and eventually culminates in the heating pattern differences previously discussed.  

\section{Conclusion}

The particle based binary algorithm's ability to capture behavior not well described by a thermal distribution and a distribution based algorithm's ability to readily use a variety of different collision rates are combined in this modification of Nanbu's cumulative collision theory. With it, one can model collisional behavior over a far wider range than pure distributional or standard binary collisions can independently, making it ideal for situations where the collisional behavior of a system changes dynamically over time and is not properly described by either model. Such scenarios include the interaction between materials and laser pulses with rapidly increasingly intensity, as well as situations where nonthermal behavior like ballistic electrons lead to back-of-the-target heating and impact ionization.  We have shown that this modification of Nanbu collisions is consistent with linear optics by modeling the absorption at low fluences and comparing it to that which is predicted by Drude theory. In addition, we show that higher fluences trigger non-thermal behavior and ballistic electrons not seen with a distributional collision model. Because of its ability to model the production and propagation of energetic particles, our algorithm can improve the modeling of laser heating for damage and surface modification studies. Finally, the algorithm described here is readily implemented in codes already using the Nanbu algorithm.

\section*{Acknowledgements}

We thank Robert Mitchell for the idea of using Colombier rates and Enam Chowdhury for useful discussions.
\newline\newline
This work was supported by the Ohio Supercomputer Center\cite{OhioSupercomputerCenter1987} and performed under the auspices of the AFOSR grants $\#$ FA9550-16-1-0069 FA9550-12-1-0454.

\bibliography{Binary_Collisions_Bibliography}

\providecommand{\noopsort}[1]{}\providecommand{\singleletter}[1]{#1}%
\begin{thebibliography}{26}%
\makeatletter
\providecommand \@ifxundefined [1]{%
 \@ifx{#1\undefined}
}%
\providecommand \@ifnum [1]{%
 \ifnum #1\expandafter \@firstoftwo
 \else \expandafter \@secondoftwo
 \fi
}%
\providecommand \@ifx [1]{%
 \ifx #1\expandafter \@firstoftwo
 \else \expandafter \@secondoftwo
 \fi
}%
\providecommand \natexlab [1]{#1}%
\providecommand \enquote  [1]{``#1''}%
\providecommand \bibnamefont  [1]{#1}%
\providecommand \bibfnamefont [1]{#1}%
\providecommand \citenamefont [1]{#1}%
\providecommand \href@noop [0]{\@secondoftwo}%
\providecommand \href [0]{\begingroup \@sanitize@url \@href}%
\providecommand \@href[1]{\@@startlink{#1}\@@href}%
\providecommand \@@href[1]{\endgroup#1\@@endlink}%
\providecommand \@sanitize@url [0]{\catcode `\\12\catcode `\$12\catcode
  `\&12\catcode `\#12\catcode `\^12\catcode `\_12\catcode `\%12\relax}%
\providecommand \@@startlink[1]{}%
\providecommand \@@endlink[0]{}%
\providecommand \url  [0]{\begingroup\@sanitize@url \@url }%
\providecommand \@url [1]{\endgroup\@href {#1}{\urlprefix }}%
\providecommand \urlprefix  [0]{URL }%
\providecommand \Eprint [0]{\href }%
\providecommand \doibase [0]{http://dx.doi.org/}%
\providecommand \selectlanguage [0]{\@gobble}%
\providecommand \bibinfo  [0]{\@secondoftwo}%
\providecommand \bibfield  [0]{\@secondoftwo}%
\providecommand \translation [1]{[#1]}%
\providecommand \BibitemOpen [0]{}%
\providecommand \bibitemStop [0]{}%
\providecommand \bibitemNoStop [0]{.\EOS\space}%
\providecommand \EOS [0]{\spacefactor3000\relax}%
\providecommand \BibitemShut  [1]{\csname bibitem#1\endcsname}%
\let\auto@bib@innerbib\@empty
\bibitem [{\citenamefont {Widmann}\ \emph {et~al.}(2004)\citenamefont
  {Widmann}, \citenamefont {Ao}, \citenamefont {Foord}, \citenamefont {Price},
  \citenamefont {Ellis}, \citenamefont {Springer},\ and\ \citenamefont
  {Ng}}]{warm_dense_matter_1}%
  \BibitemOpen
  \bibfield  {author} {\bibinfo {author} {\bibfnamefont {K.}~\bibnamefont
  {Widmann}}, \bibinfo {author} {\bibfnamefont {T.}~\bibnamefont {Ao}},
  \bibinfo {author} {\bibfnamefont {M.~E.}\ \bibnamefont {Foord}}, \bibinfo
  {author} {\bibfnamefont {D.~F.}\ \bibnamefont {Price}}, \bibinfo {author}
  {\bibfnamefont {A.~D.}\ \bibnamefont {Ellis}}, \bibinfo {author}
  {\bibfnamefont {P.~T.}\ \bibnamefont {Springer}}, \ and\ \bibinfo {author}
  {\bibfnamefont {A.}~\bibnamefont {Ng}},\ }\href@noop {} {\bibfield  {journal}
  {\bibinfo  {journal} {Phys.\ Rev.\ Lett.}\ }\textbf {\bibinfo {volume}
  {92}},\ \bibinfo {pages} {125002} (\bibinfo {year} {2004})}\BibitemShut
  {NoStop}%
\bibitem [{\citenamefont {More}\ \emph {et~al.}(2006)\citenamefont {More},
  \citenamefont {Yoneda},\ and\ \citenamefont
  {Morikami}}]{warm_dense_matter_2}%
  \BibitemOpen
  \bibfield  {author} {\bibinfo {author} {\bibfnamefont {R.}~\bibnamefont
  {More}}, \bibinfo {author} {\bibfnamefont {H.}~\bibnamefont {Yoneda}}, \ and\
  \bibinfo {author} {\bibfnamefont {H.}~\bibnamefont {Morikami}},\ }\href@noop
  {} {\bibfield  {journal} {\bibinfo  {journal} {Journal of Quantitative
  Spectroscopy and Radiative Transfer}\ }\textbf {\bibinfo {volume} {99}},\
  \bibinfo {pages} {409} (\bibinfo {year} {2006})}\BibitemShut {NoStop}%
\bibitem [{\citenamefont {Gakovic}\ \emph {et~al.}(2009)\citenamefont
  {Gakovic}, \citenamefont {Stasic}, \citenamefont {Petrovic}, \citenamefont
  {Radak}, \citenamefont {Krmpot}, \citenamefont {Jelenkovic},\ and\
  \citenamefont {Trtica}}]{surface_modification_1}%
  \BibitemOpen
  \bibfield  {author} {\bibinfo {author} {\bibfnamefont {B.}~\bibnamefont
  {Gakovic}}, \bibinfo {author} {\bibfnamefont {J.}~\bibnamefont {Stasic}},
  \bibinfo {author} {\bibfnamefont {S.}~\bibnamefont {Petrovic}}, \bibinfo
  {author} {\bibfnamefont {B.}~\bibnamefont {Radak}}, \bibinfo {author}
  {\bibfnamefont {A.}~\bibnamefont {Krmpot}}, \bibinfo {author} {\bibfnamefont
  {B.}~\bibnamefont {Jelenkovic}}, \ and\ \bibinfo {author} {\bibfnamefont
  {M.}~\bibnamefont {Trtica}},\ }\href@noop {} {\bibfield  {journal} {\bibinfo
  {journal} {Acta Physica Polonica A}\ }\textbf {\bibinfo {volume} {116}},\
  \bibinfo {pages} {611} (\bibinfo {year} {2009})}\BibitemShut {NoStop}%
\bibitem [{\citenamefont {M{\'e}zel}\ \emph {et~al.}(2010)\citenamefont
  {M{\'e}zel}, \citenamefont {Bourgeade},\ and\ \citenamefont
  {Hallo}}]{surface_modification_2}%
  \BibitemOpen
  \bibfield  {author} {\bibinfo {author} {\bibfnamefont {C.}~\bibnamefont
  {M{\'e}zel}}, \bibinfo {author} {\bibfnamefont {A.}~\bibnamefont
  {Bourgeade}}, \ and\ \bibinfo {author} {\bibfnamefont {L.}~\bibnamefont
  {Hallo}},\ }\href@noop {} {\bibfield  {journal} {\bibinfo  {journal} {Phys.
  Plasmas}\ }\textbf {\bibinfo {volume} {17}},\ \bibinfo {pages} {113504}
  (\bibinfo {year} {2010})}\BibitemShut {NoStop}%
\bibitem [{\citenamefont {Stuart}\ \emph {et~al.}(1996)\citenamefont {Stuart},
  \citenamefont {Feit}, \citenamefont {Herman}, \citenamefont {Rubenchik},
  \citenamefont {Shore},\ and\ \citenamefont {Perry}}]{optical_ablation}%
  \BibitemOpen
  \bibfield  {author} {\bibinfo {author} {\bibfnamefont {B.~C.}\ \bibnamefont
  {Stuart}}, \bibinfo {author} {\bibfnamefont {M.~D.}\ \bibnamefont {Feit}},
  \bibinfo {author} {\bibfnamefont {S.}~\bibnamefont {Herman}}, \bibinfo
  {author} {\bibfnamefont {A.~M.}\ \bibnamefont {Rubenchik}}, \bibinfo {author}
  {\bibfnamefont {B.~W.}\ \bibnamefont {Shore}}, \ and\ \bibinfo {author}
  {\bibfnamefont {M.~D.}\ \bibnamefont {Perry}},\ }\href@noop {} {\bibfield
  {journal} {\bibinfo  {journal} {JOSA B}\ }\textbf {\bibinfo {volume} {13}},\
  \bibinfo {pages} {459} (\bibinfo {year} {1996})}\BibitemShut {NoStop}%
\bibitem [{\citenamefont {Mitchell}\ \emph {et~al.}(2015)\citenamefont
  {Mitchell}, \citenamefont {Schumacher},\ and\ \citenamefont
  {Chowdhury}}]{Rob}%
  \BibitemOpen
  \bibfield  {author} {\bibinfo {author} {\bibfnamefont {R.~A.}\ \bibnamefont
  {Mitchell}}, \bibinfo {author} {\bibfnamefont {D.~W.}\ \bibnamefont
  {Schumacher}}, \ and\ \bibinfo {author} {\bibfnamefont {E.~A.}\ \bibnamefont
  {Chowdhury}},\ }\href@noop {} {\bibfield  {journal} {\bibinfo  {journal}
  {Optics Letters}\ }\textbf {\bibinfo {volume} {40}},\ \bibinfo {pages} {2189}
  (\bibinfo {year} {2015})}\BibitemShut {NoStop}%
\bibitem [{\citenamefont {Plamann}\ \emph {et~al.}(2010)\citenamefont
  {Plamann}, \citenamefont {Aptel}, \citenamefont {Arnold}, \citenamefont
  {Courjaud}, \citenamefont {Crotti}, \citenamefont {Deloison}, \citenamefont
  {Druon}, \citenamefont {Georges}, \citenamefont {Hanna}, \citenamefont
  {Legeais}, \citenamefont {Morin}, \citenamefont {Mottay}, \citenamefont
  {Peyrot},\ and\ \citenamefont {Savoldelli}}]{eye_surgery}%
  \BibitemOpen
  \bibfield  {author} {\bibinfo {author} {\bibfnamefont {K.}~\bibnamefont
  {Plamann}}, \bibinfo {author} {\bibfnamefont {F.}~\bibnamefont {Aptel}},
  \bibinfo {author} {\bibfnamefont {C.~L.}\ \bibnamefont {Arnold}}, \bibinfo
  {author} {\bibfnamefont {A.}~\bibnamefont {Courjaud}}, \bibinfo {author}
  {\bibfnamefont {C.}~\bibnamefont {Crotti}}, \bibinfo {author} {\bibfnamefont
  {F.}~\bibnamefont {Deloison}}, \bibinfo {author} {\bibfnamefont
  {F.}~\bibnamefont {Druon}}, \bibinfo {author} {\bibfnamefont
  {P.}~\bibnamefont {Georges}}, \bibinfo {author} {\bibfnamefont
  {M.}~\bibnamefont {Hanna}}, \bibinfo {author} {\bibfnamefont {J.-M.}\
  \bibnamefont {Legeais}}, \bibinfo {author} {\bibfnamefont {F.}~\bibnamefont
  {Morin}}, \bibinfo {author} {\bibfnamefont {E.}~\bibnamefont {Mottay}},
  \bibinfo {author} {\bibfnamefont {V.~N. D.~A.}\ \bibnamefont {Peyrot}}, \
  and\ \bibinfo {author} {\bibfnamefont {M.}~\bibnamefont {Savoldelli}},\
  }\href@noop {} {\bibfield  {journal} {\bibinfo  {journal} {J. Opt.}\ }\textbf
  {\bibinfo {volume} {12}},\ \bibinfo {pages} {084002} (\bibinfo {year}
  {2010})}\BibitemShut {NoStop}%
\bibitem [{\citenamefont {Liu}\ and\ \citenamefont
  {Mourou}(1997)}]{laser_machining_1}%
  \BibitemOpen
  \bibfield  {author} {\bibinfo {author} {\bibfnamefont {X.}~\bibnamefont
  {Liu}}\ and\ \bibinfo {author} {\bibfnamefont {G.}~\bibnamefont {Mourou}},\
  }\href@noop {} {\bibfield  {journal} {\bibinfo  {journal} {Laser Focus
  World}\ }\textbf {\bibinfo {volume} {33}},\ \bibinfo {pages} {101} (\bibinfo
  {year} {1997})}\BibitemShut {NoStop}%
\bibitem [{\citenamefont {Perry}\ \emph {et~al.}(1999)\citenamefont {Perry},
  \citenamefont {Stuart}, \citenamefont {Banks}, \citenamefont {Feit},
  \citenamefont {Yanovsky},\ and\ \citenamefont
  {Rubenchik}}]{laser_machining_2}%
  \BibitemOpen
  \bibfield  {author} {\bibinfo {author} {\bibfnamefont {M.~D.}\ \bibnamefont
  {Perry}}, \bibinfo {author} {\bibfnamefont {B.~C.}\ \bibnamefont {Stuart}},
  \bibinfo {author} {\bibfnamefont {P.~S.}\ \bibnamefont {Banks}}, \bibinfo
  {author} {\bibfnamefont {M.~D.}\ \bibnamefont {Feit}}, \bibinfo {author}
  {\bibfnamefont {V.}~\bibnamefont {Yanovsky}}, \ and\ \bibinfo {author}
  {\bibfnamefont {A.~M.}\ \bibnamefont {Rubenchik}},\ }\href@noop {} {\bibfield
   {journal} {\bibinfo  {journal} {J. Appl. Phys.}\ }\textbf {\bibinfo {volume}
  {85}},\ \bibinfo {pages} {6803} (\bibinfo {year} {1999})}\BibitemShut
  {NoStop}%
\bibitem [{\citenamefont {Wu}\ \emph {et~al.}(2011)\citenamefont {Wu},
  \citenamefont {Cheng}, \citenamefont {Chang}, \citenamefont {Wu},\ and\
  \citenamefont {Wang}}]{hydrophobic_1}%
  \BibitemOpen
  \bibfield  {author} {\bibinfo {author} {\bibfnamefont {P.~H.}\ \bibnamefont
  {Wu}}, \bibinfo {author} {\bibfnamefont {C.~W.}\ \bibnamefont {Cheng}},
  \bibinfo {author} {\bibfnamefont {C.~P.}\ \bibnamefont {Chang}}, \bibinfo
  {author} {\bibfnamefont {T.~M.}\ \bibnamefont {Wu}}, \ and\ \bibinfo {author}
  {\bibfnamefont {J.~K.}\ \bibnamefont {Wang}},\ }\href@noop {} {\bibfield
  {journal} {\bibinfo  {journal} {J. Micromech, Microeng.}\ }\textbf {\bibinfo
  {volume} {21}},\ \bibinfo {pages} {115032} (\bibinfo {year}
  {2011})}\BibitemShut {NoStop}%
\bibitem [{\citenamefont {Cardoso}\ \emph {et~al.}(2011)\citenamefont
  {Cardoso}, \citenamefont {Tribuzi}, \citenamefont {Balogh}, \citenamefont
  {Misoguti},\ and\ \citenamefont {Mendonca}}]{hydrophobic_2}%
  \BibitemOpen
  \bibfield  {author} {\bibinfo {author} {\bibfnamefont {M.~R.}\ \bibnamefont
  {Cardoso}}, \bibinfo {author} {\bibfnamefont {V.}~\bibnamefont {Tribuzi}},
  \bibinfo {author} {\bibfnamefont {D.~T.}\ \bibnamefont {Balogh}}, \bibinfo
  {author} {\bibfnamefont {L.}~\bibnamefont {Misoguti}}, \ and\ \bibinfo
  {author} {\bibfnamefont {C.~R.}\ \bibnamefont {Mendonca}},\ }\href@noop {}
  {\bibfield  {journal} {\bibinfo  {journal} {Applied Surface Science}\
  }\textbf {\bibinfo {volume} {257}},\ \bibinfo {pages} {3281} (\bibinfo {year}
  {2011})}\BibitemShut {NoStop}%
\bibitem [{\citenamefont {Kim}\ \emph {et~al.}(2013)\citenamefont {Kim},
  \citenamefont {Ki},\ and\ \citenamefont {Park}}]{solar_cell}%
  \BibitemOpen
  \bibfield  {author} {\bibinfo {author} {\bibfnamefont {K.~R.}\ \bibnamefont
  {Kim}}, \bibinfo {author} {\bibfnamefont {T.~H.}\ \bibnamefont {Ki}}, \ and\
  \bibinfo {author} {\bibfnamefont {H.~A.}\ \bibnamefont {Park}},\ }\href@noop
  {} {\bibfield  {journal} {\bibinfo  {journal} {Applied Surface Science}\
  }\textbf {\bibinfo {volume} {264}},\ \bibinfo {pages} {404} (\bibinfo {year}
  {2013})}\BibitemShut {NoStop}%
\bibitem [{\citenamefont {Mizoshiri}\ \emph {et~al.}(2010)\citenamefont
  {Mizoshiri}, \citenamefont {Nishiyama}, \citenamefont {Nishii},\ and\
  \citenamefont {Hirata}}]{lithography}%
  \BibitemOpen
  \bibfield  {author} {\bibinfo {author} {\bibfnamefont {M.}~\bibnamefont
  {Mizoshiri}}, \bibinfo {author} {\bibfnamefont {H.}~\bibnamefont
  {Nishiyama}}, \bibinfo {author} {\bibfnamefont {J.}~\bibnamefont {Nishii}}, \
  and\ \bibinfo {author} {\bibfnamefont {Y.}~\bibnamefont {Hirata}},\
  }\href@noop {} {\bibfield  {journal} {\bibinfo  {journal} {Appl. Phys. A.}\
  }\textbf {\bibinfo {volume} {98}},\ \bibinfo {pages} {171} (\bibinfo {year}
  {2010})}\BibitemShut {NoStop}%
\bibitem [{\citenamefont {Takizuka}\ and\ \citenamefont
  {Abe}(1977)}]{Takizuka}%
  \BibitemOpen
  \bibfield  {author} {\bibinfo {author} {\bibfnamefont {T.}~\bibnamefont
  {Takizuka}}\ and\ \bibinfo {author} {\bibfnamefont {H.}~\bibnamefont {Abe}},\
  }\href@noop {} {\bibfield  {journal} {\bibinfo  {journal} {J. Comput. Phys.}\
  }\textbf {\bibinfo {volume} {25}},\ \bibinfo {pages} {205} (\bibinfo {year}
  {1977})}\BibitemShut {NoStop}%
\bibitem [{\citenamefont {Nanbu}(1997)}]{Nanbu}%
  \BibitemOpen
  \bibfield  {author} {\bibinfo {author} {\bibfnamefont {K.}~\bibnamefont
  {Nanbu}},\ }\href@noop {} {\bibfield  {journal} {\bibinfo  {journal} {Phys.
  Rev. E.}\ }\textbf {\bibinfo {volume} {55}},\ \bibinfo {pages} {4642}
  (\bibinfo {year} {1997})}\BibitemShut {NoStop}%
\bibitem [{\citenamefont {L.{ Spitzer Jr.}}(1967)}]{Spitzer}%
  \BibitemOpen
  \bibfield  {author} {\bibinfo {author} {\bibnamefont {L.{ Spitzer Jr.}}},\
  }\href@noop {} {\emph {\bibinfo {title} {Physics of Fully Ionized Gases}}},\
  \bibinfo {edition} {2nd}\ ed.\ (\bibinfo  {publisher} {Interscience},\
  \bibinfo {address} {New York},\ \bibinfo {year} {1967})\BibitemShut {NoStop}%
\bibitem [{\citenamefont {Wagner}\ \emph {et~al.}(2016)\citenamefont {Wagner},
  \citenamefont {Deppert}, \citenamefont {Brabetz}, \citenamefont {Fiala},
  \citenamefont {Kleinschmidt}, \citenamefont {Poth}, \citenamefont {Schanz},
  \citenamefont {Tebartz}, \citenamefont {Zielbauer}, \citenamefont {Roth},
  \citenamefont {St{\"o}hlker},\ and\ \citenamefont {Bagnoud}}]{GSI}%
  \BibitemOpen
  \bibfield  {author} {\bibinfo {author} {\bibfnamefont {F.}~\bibnamefont
  {Wagner}}, \bibinfo {author} {\bibfnamefont {O.}~\bibnamefont {Deppert}},
  \bibinfo {author} {\bibfnamefont {C.}~\bibnamefont {Brabetz}}, \bibinfo
  {author} {\bibfnamefont {P.}~\bibnamefont {Fiala}}, \bibinfo {author}
  {\bibfnamefont {A.}~\bibnamefont {Kleinschmidt}}, \bibinfo {author}
  {\bibfnamefont {P.}~\bibnamefont {Poth}}, \bibinfo {author} {\bibfnamefont
  {V.~A.}\ \bibnamefont {Schanz}}, \bibinfo {author} {\bibfnamefont
  {A.}~\bibnamefont {Tebartz}}, \bibinfo {author} {\bibfnamefont
  {B.}~\bibnamefont {Zielbauer}}, \bibinfo {author} {\bibfnamefont
  {M.}~\bibnamefont {Roth}}, \bibinfo {author} {\bibfnamefont {T.}~\bibnamefont
  {St{\"o}hlker}}, \ and\ \bibinfo {author} {\bibfnamefont {V.}~\bibnamefont
  {Bagnoud}},\ }\href@noop {} {\bibfield  {journal} {\bibinfo  {journal}
  {Phys.\ Rev.\ Lett.}\ }\textbf {\bibinfo {volume} {116}},\ \bibinfo {pages}
  {205002} (\bibinfo {year} {2016})}\BibitemShut {NoStop}%
\bibitem [{\citenamefont {Welch}\ \emph {et~al.}(2006)\citenamefont {Welch},
  \citenamefont {Rose}, \citenamefont {Cuneo}, \citenamefont {Campbell},\ and\
  \citenamefont {Mehlhorn}}]{LSP}%
  \BibitemOpen
  \bibfield  {author} {\bibinfo {author} {\bibfnamefont {D.~R.}\ \bibnamefont
  {Welch}}, \bibinfo {author} {\bibfnamefont {D.~V.}\ \bibnamefont {Rose}},
  \bibinfo {author} {\bibfnamefont {M.~E.}\ \bibnamefont {Cuneo}}, \bibinfo
  {author} {\bibfnamefont {R.~B.}\ \bibnamefont {Campbell}}, \ and\ \bibinfo
  {author} {\bibfnamefont {T.~A.}\ \bibnamefont {Mehlhorn}},\ }\href@noop {}
  {\bibfield  {journal} {\bibinfo  {journal} {Phys. Plasmas}\ }\textbf
  {\bibinfo {volume} {13}},\ \bibinfo {pages} {063105} (\bibinfo {year}
  {2006})}\BibitemShut {NoStop}%
\bibitem [{\citenamefont {Jackson}(1999)}]{Jackson}%
  \BibitemOpen
  \bibfield  {author} {\bibinfo {author} {\bibfnamefont {J.~D.}\ \bibnamefont
  {Jackson}},\ }\href@noop {} {\emph {\bibinfo {title} {Classical
  Electrodynamics}}},\ \bibinfo {edition} {3rd}\ ed.\ (\bibinfo  {publisher}
  {John Wiley and Sons, Inc.},\ \bibinfo {address} {New Jersey},\ \bibinfo
  {year} {1999})\BibitemShut {NoStop}%
\bibitem [{\citenamefont {Wang}\ \emph {et~al.}(2008)\citenamefont {Wang},
  \citenamefont {Lin}, \citenamefont {Caflisch}, \citenamefont {Cohen},\ and\
  \citenamefont {Dimits}}]{Nanbu_vs_TA}%
  \BibitemOpen
  \bibfield  {author} {\bibinfo {author} {\bibfnamefont {C.}~\bibnamefont
  {Wang}}, \bibinfo {author} {\bibfnamefont {T.}~\bibnamefont {Lin}}, \bibinfo
  {author} {\bibfnamefont {R.}~\bibnamefont {Caflisch}}, \bibinfo {author}
  {\bibfnamefont {B.}~\bibnamefont {Cohen}}, \ and\ \bibinfo {author}
  {\bibfnamefont {A.~M.}\ \bibnamefont {Dimits}},\ }\href@noop {} {\bibfield
  {journal} {\bibinfo  {journal} {J. Comp. Phys.}\ }\textbf {\bibinfo {volume}
  {227}},\ \bibinfo {pages} {4308} (\bibinfo {year} {2008})}\BibitemShut
  {NoStop}%
\bibitem [{\citenamefont {Lee}\ and\ \citenamefont {More}(1984)}]{LM_Rates}%
  \BibitemOpen
  \bibfield  {author} {\bibinfo {author} {\bibfnamefont {Y.~T.}\ \bibnamefont
  {Lee}}\ and\ \bibinfo {author} {\bibfnamefont {R.~M.}\ \bibnamefont {More}},\
  }\href@noop {} {\bibfield  {journal} {\bibinfo  {journal} {Phys. Fluids}\
  }\textbf {\bibinfo {volume} {27}},\ \bibinfo {pages} {1273} (\bibinfo {year}
  {1984})}\BibitemShut {NoStop}%
\bibitem [{\citenamefont {Desjarlais}(2001)}]{LMD_Rates}%
  \BibitemOpen
  \bibfield  {author} {\bibinfo {author} {\bibfnamefont {M.~P.}\ \bibnamefont
  {Desjarlais}},\ }\href@noop {} {\bibfield  {journal} {\bibinfo  {journal}
  {Contrib. Plasma Phys.}\ }\textbf {\bibinfo {volume} {41}},\ \bibinfo {pages}
  {267} (\bibinfo {year} {2001})}\BibitemShut {NoStop}%
\bibitem [{\citenamefont {Colombier}\ \emph {et~al.}(2008)\citenamefont
  {Colombier}, \citenamefont {Combis}, \citenamefont {Audouard},\ and\
  \citenamefont {Stoian}}]{colombier}%
  \BibitemOpen
  \bibfield  {author} {\bibinfo {author} {\bibfnamefont {J.~P.}\ \bibnamefont
  {Colombier}}, \bibinfo {author} {\bibfnamefont {P.}~\bibnamefont {Combis}},
  \bibinfo {author} {\bibfnamefont {E.}~\bibnamefont {Audouard}}, \ and\
  \bibinfo {author} {\bibfnamefont {R.}~\bibnamefont {Stoian}},\ }\href@noop {}
  {\bibfield  {journal} {\bibinfo  {journal} {Phys. Rev. E}\ }\textbf {\bibinfo
  {volume} {77}},\ \bibinfo {pages} {036409} (\bibinfo {year}
  {2008})}\BibitemShut {NoStop}%
\bibitem [{\citenamefont {Jones}\ \emph {et~al.}(1995)\citenamefont {Jones},
  \citenamefont {Lemons}, \citenamefont {Mason}, \citenamefont {Thomas},\ and\
  \citenamefont {Winske}}]{Jones}%
  \BibitemOpen
  \bibfield  {author} {\bibinfo {author} {\bibfnamefont {M.~E.}\ \bibnamefont
  {Jones}}, \bibinfo {author} {\bibfnamefont {D.~S.}\ \bibnamefont {Lemons}},
  \bibinfo {author} {\bibfnamefont {R.~J.}\ \bibnamefont {Mason}}, \bibinfo
  {author} {\bibfnamefont {V.~A.}\ \bibnamefont {Thomas}}, \ and\ \bibinfo
  {author} {\bibfnamefont {D.}~\bibnamefont {Winske}},\ }\href@noop {}
  {\bibfield  {journal} {\bibinfo  {journal} {J. Comput. Phys.}\ }\textbf
  {\bibinfo {volume} {123}},\ \bibinfo {pages} {169} (\bibinfo {year}
  {1995})}\BibitemShut {NoStop}%
\bibitem [{\citenamefont {Ashcroft}\ and\ \citenamefont
  {Mermin}(1976)}]{Drude}%
  \BibitemOpen
  \bibfield  {author} {\bibinfo {author} {\bibfnamefont {N.}~\bibnamefont
  {Ashcroft}}\ and\ \bibinfo {author} {\bibfnamefont {N.~D.}\ \bibnamefont
  {Mermin}},\ }\href@noop {} {\emph {\bibinfo {title} {Solid State Physics}}},\
  \bibinfo {edition} {1st}\ ed.\ (\bibinfo  {publisher} {Brooks/Cole},\
  \bibinfo {address} {California},\ \bibinfo {year} {1976})\BibitemShut
  {NoStop}%
\bibitem [{\citenamefont {Center}(1987)}]{OhioSupercomputerCenter1987}%
  \BibitemOpen
  \bibfield  {author} {\bibinfo {author} {\bibfnamefont {O.~S.}\ \bibnamefont
  {Center}},\ }\href@noop {} {\enquote {\bibinfo {title} {Ohio supercomputer
  center},}\ }\bibinfo {howpublished}
  {\url{http://osc.edu/ark:/19495/f5s1ph73}} (\bibinfo {year}
  {1987})\BibitemShut {NoStop}%
\end{thebibliography}%
\end{document}